\documentclass[aps,floatfix,superscriptaddress,notitlepage]{revtex4-1}
\usepackage{amsmath,amssymb,amsfonts,graphics,graphicx,dcolumn,bm,enumerate}
\usepackage{comment,natbib,appendix}
\usepackage{multirow,color}
\usepackage{chngpage}
\usepackage{afterpage}
\usepackage{xcolor}
\usepackage{amsthm}
\usepackage{epsfig}
\usepackage{natbib}
\usepackage{hyperref}
\usepackage{graphicx}
\usepackage{soul}

\newcommand{\cmp}
{\affiliation{Saha Institute of Nuclear Physics, Kolkata 700064, India.}}

\newcommand{\isi}
{\affiliation{Economic Research Unit, Indian Statistical Institute, Kolkata 700108, India.}}

\newcommand{\raghunathpur}
{\affiliation{Department of Physics, Raghunathpur College, Raghunathpur, Purulia 723133, India.}}

\newcommand{\snb}
{\affiliation{Satyendra Nath Bose National Centre for Basic
Sciences, Block-JD, Salt Lake, Kolkata-700106, India.}}

\newcommand{\Barasat}
{\affiliation{Department of Physics, Barasat Government College, Kolkata 700124, India.}}

\newcommand{\Xaviers}
{\affiliation{St. Xaviers College, Mumbai 400001, India.}}

\begin{document}

\title{Kinetic Exchange Income Distribution  Models with Saving Propensities: Inequality Indices and  Self-Organised Poverty Level}

\author{Sanjukta Paul}
 \email[Email: ]{sanjuktapaul024@gmail.com}
 \snb
 
\author{Sudip Mukherjee}
 \email[Email: ]{sudip.bat@gmail.com}
 \Barasat \cmp
 
 \author{Bijin Joseph}
 \email[Email: ]{bijin.joseph@xaviers.edu.in}
 \Xaviers
 
 \author{Asim Ghosh}
 \email[Email: ]{asimghosh066@gmail.com}
 \raghunathpur
 
 \author{Bikas K. Chakrabarti}
 \email[Email: ]{bikask.chakrabarti@saha.ac.in}
\snb\cmp \isi

\begin{abstract}
We report  the numerical results for the steady state income or wealth distribution $P(m)$ and the resulting inequality measures (Gini $g$ and Kolkata $k$ indices) in the kinetic exchange models of market dynamics. We study the variations of $P(m)$
and of the indices $g$ and $k$ with the saving
propensity $\lambda$ of the agents, with two
different kinds of trade (kinetic exchange) dynamics.
In the first case, the exchange occurs  between
randomly chosen pairs of agents
and in the next, one
of the agents in the chosen pair is the poorest of
all and the other agent is randomly picked up from
the rest of the population (where, in the steady state, a self-organized
poverty level or SOPL appears).  These studies
have also been made for two different kinds of saving
behaviors. One, where each agent has the same value of
$\lambda$  (constant over time) and the other where
$\lambda$ for each agent can take two values (0 and
1), changing randomly over a fraction
of time $\rho(<1)$ of choosing $\lambda = 1$. 
We find that the inequality decreases with increasing
savings ($\lambda$); inequality indices ($g$ and $k$) decrease and SOPL increases with increasing $\lambda$,
indicating possible applications in economic policy making.
\end{abstract}



\maketitle
\section{Introduction}
The kinetic theory of gases, more
than a century old and the first successful
classical many-body theory of condensed matter physics,  has  recently been applied in
econophysics and sociophysics  (see e.g., \cite{sinha2010econophysics,chakrabarti2006econophysics})
in the  modeling of different  socio-economic contexts. 
Statistically modelling the market of traders to be a thermodynamic system in equilibrium with a 
large number of interacting gas molecules, the uneven distribution of wealth of individuals trading in the market 
can be estimated using the idea of the energy distribution of the gas molecules obtained using kinetic theory of gases.
These two-body exchange dynamics studies have thus been
extensively developed in the context of modeling
income or wealth distributions in a society (see
e.g., \cite{yakovenko2009colloquium,chakrabarti2013econophysics,pareschi2013interacting,ribeiro2020income}). For extensions of these kinetic
exchange models in the case of social opinion formation
studies, see e.g., \cite{sen2014sociophysics,galam2012sociophysics,pareschi2013interacting}.

In generic kinetic exchange models of income or wealth distributions in a society, one studies a system of $N$ agents who interact among themselves through two-agent money ($m$) conserving stochastic trade  (scattering) processes, where each agent saves a fraction $\lambda$ of the money or wealth at each trade (exchange) or instant of time \cite{chakraborti2000statistical,chatterjee2004pareto}. 
While a model with no saving behavior for the agents yields the well known Gibbs distribution for the steady state, a non-zero saving propensity results in the market being interacting, with steady state wealth distributions having a most probable value and exponentially decaying on both sides of it.
The resulting steady state distributions $P(m)$ of money, for different values of the saving propensities $\lambda$ are compared with the available data (see e.g., \cite{chakrabarti2013econophysics,chatterjee2007kinetic}). 
While a peaked distribution (better fitted to a Gamma distribution) arises simply as a result of individual saving tendencies of the population, a more involved feature in the market economy like a self generated poverty level can arise through selective trading mechanisms.
In this regard, one can also study the effect of modification in the kinetic exchange dynamics such that one of the agents
in any chosen pair participating in the
trading (or scattering) process has the
lowest money or wealth at
that point of time, while the other agent is
selected randomly from the rest of the population,
with no constraint on the amount of money possessed. \cite{pianegonda2003wealth,ghosh2011threshold}. 
Alternatively, one can also choose the pair of agents based on their total wealth, such that
one of them has money below an arbitrarily chosen poverty-level and the other one,
selected randomly from the whole population can have any amount of money. The
kinetic exchange dynamics is continued until no agent is left with money below a new
Self-Organized Poverty-Line or SOPL \cite{ghosh2011threshold} (see also \cite{iglesias2010simple,chakrabarti2021development}). Then by varying $\lambda$,
it is investigated whether the SOPL can be shifted to higher values of money. 
The motivation for our study is threefold -  (i) to generate a clear idea
about whether and under what conditions can the well established $\lambda = 0$ 
exponential, with $m = 0$ recording the highest $P(m)$, be shifted to a non-zero
$m$ value such that most of the agents are not paupers, (ii) to reduce the social
inequalities by reducing the inequality indices g and k and (iii) to calculate SOPL 
as a function of $\lambda$ (fixed over time) and $\rho$ ($\lambda$ varying over time) with the
aim of finding the condition for which it increases, so that higher 
number of agents participating in the trading exchange can have wealth at least
above a threshold value of $m$, pointing to a more stable economy.

The resulting inequalities can be measured here by
determining  the Most Probable Income (MPI), given
by the location of the maximum value of the distribution
$P(m)$, or by the location of the SOPL (below which
$P(m) = 0$) together with the determination of the
values of the Gini ($g$) and Kolkata ($k$) indices (see
e.g., \cite{chakrabarti2021development,banerjee2020inequality}). 

The rest of the paper progresses as follows. We numerically study the variations in income or wealth distribution $P(m)$ generated from the kinetic exchange models described below (section \ref{results}) and extract the variations in the values of Gini $g$ and Kolkata $k$ indices (defined in section \ref{theory}) with the saving propensity $\lambda$ of the agents.  We consider two different kinds of trade or (kinetic) exchange dynamics (see e.g. \cite{sinha2020econophysics} for more information regarding the simulations of kinetic exchange models) with two different saving behaviors as described in subsections \ref{cc-model}, \ref{soc-model}, \ref{cc-two} and \ref{soc-two} . In section \ref{conclusions} we summarize
our work and conclude with  discussions.

\section{Theoretical background for the inequality indices}
\label{theory}
\begin{figure}[!tbh]
    \centering
    \includegraphics[width=0.7\textwidth]{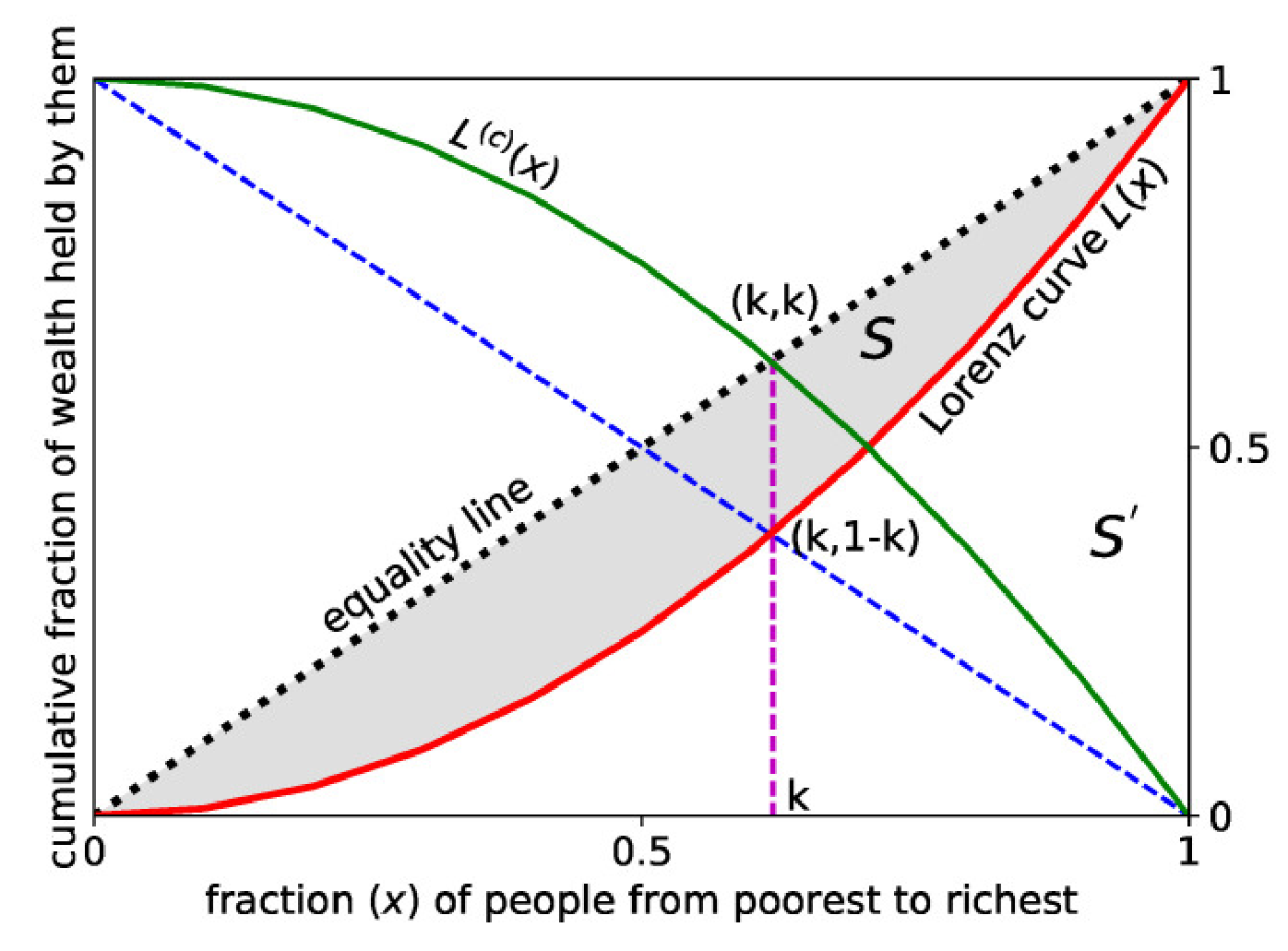}
    \caption{The Lorenz function $L(x)$ represents the
fraction (curve shown in red) of overall income or wealth assumed by the
poorer $x$ fraction of the people. The Gini
index $g$ is given by the ratio of the area that lies
between the line of equality and Lorenz curve over
the total area under the line of equality ($g = 2S$). The
complementary Lorenz function $L^c(x) \equiv 1-L(x)$
is represented here by the green curve. The $k$
index value is given by the fixed point $L^c(k) = k$
of $L^c(x)$.}
    \label{fig:1}
\end{figure}

Both the indices, Gini (oldest and most popular
one) and Kolkata (introduced in \cite{ghosh2014inequality}, see \cite{banerjee2020inequality}
for a recent review and \cite{west2021crucial} for a recent
discussion in the context of empirical studies on
'Science of Science'), are based on the Lorenz
curve or function (see \cite{chakrabarti2021development,banerjee2020inequality}) $L(x)$, giving
the cumulative fraction ($L = \int_0^m m P(m)dm$/
$[\int_0^{\infty} m P(m)dm]$) of (total accumulated)
income or wealth possessed by  the fraction
($x = \int_0^m  P(m)dm/[\int_0^{\infty} P(m)dm]$)
of the population, when counted from the poorest to
the richest (see Fig. \ref{fig:1}).

If the income (wealth) of every agent
is identical, then $L(x)$ will be a linear function
represented by the diagonal (in Fig. \ref{fig:1}) passing
through the origin. This diagonal defined by $L(x) = x$,   is called the
equality line. The Gini coefficient ($g$) is
given by the area between the Lorenz curve $L(x)$
and the equality line (the normalized area
under the equality line): $g$ = 0 corresponds to
equality and $g$ = 1 corresponds to extreme
inequality. The Kolkata index or $k$-index is
given by the ordinate value $k$ of the intersecting
point of the Lorenz curve and the diagonal
perpendicular to the equality line. By construction
(see Fig. \ref{fig:1}) $L(k) = 1-k$, and  that $k$ fraction
of wealth is  possessed by ($1-k$) fraction of
the richest population. As such, it gives a
quantitative generalization of the approximately
established (phenomenological) 80-20
law of Pareto (see e.g., \cite{banerjee2020inequality}), indicating that
typically about $80\%$ wealth is possessed
by only $20\%$ of the richest population in any economy. Defining the
complementary Lorenz function $L^c(x) \equiv
[1 - L(x)]$, one gets $k$ as its (nontrivial) fixed point (while Lorenz function $L(x)$
itself has trivial fixed points at $x$ = 0 and 1). Note, $k$ = 0.5 corresponds 
to complete equality and $k$ = 1 corresponds to extreme inequality.
As an example, both $g$ and $k$ may be exactly calculated
or estimated in the well known case of Gibbs
distribution (normalized) $P(m) = \exp(-m)$: With
$x =\int_0^m \exp (-m')dm' = 1- \exp(-m)$, giving
$m = - \ln(1-x)$, and $L = \int_0^m m'\exp(-m')dm'$
= $1 - (m + 1)\exp(-m)$, giving $L(x) = $
$1 - (1 - x)[1- \ln(1 - x)]$. As the area under the
equality line is 1/2, the Gini index $g = $
$1-2\int_0^1 L(x)dx = 1/2$ and the Kolkata  index $k$
for this Gibbs (exponential) distribution is given
by the self-consistent equation $1 - k = L(k)$ or
$1 - 2k = (1-k)[\ln(1 - k)]$, giving $k \simeq 0.68$.


\section{Models and Simulation Results}
\label{results}
In this section, we will discuss the numerical
studies of the kinetic exchange models of income
distribution (among agents or traders having
saving propensities $\lambda$), employing two
kinds of dynamics. In subsection \ref{cc-model}, the exchange occurs between randomly chosen pairs of agents with uniform saving behavior $0 \leq \lambda < 1$, which is fixed over time (here the most probable income in the steady-state increases with increasing saving propensity $\lambda$; see \cite{chakraborti2000statistical}). In \ref{soc-model}, one of the agents or traders in the chosen pair is the poorest in the whole population at that point of time (trade, exchange, or scattering) while the other one is randomly picked up from the rest, with the agents having a uniform saving propensity $0 \leq \lambda < 1$, fixed over time (where, in the steady-state, a self-organized minimum poverty level or SOPL of income or wealth appears, see \cite{ghosh2011threshold} for agents with saving propensity $\lambda=0$). Here we show that SOPL increases with increasing $\lambda$. In subsection \ref{cc-two}, we revisit the trade dynamics where the exchange occurs between randomly chosen pairs of agents with the modification that $\lambda$ for each agent can take two values, zero (with probability $1-\rho$, $\rho <1$) or unity (with probability $\rho$), and it changes randomly over time or trade. 
Finally, in subsection \ref{soc-two} 
the saving behaviour followed is that described in subsection \ref{cc-two}
but with the trading dynamics described in subsection \ref{soc-model}, the one that produces a self-organized poverty level.

We perform numerical  simulations with fixed number of agents $N$ and total money $M=N$ for  both the models. In our simulation, one Monte Carlo step is equivalent to $N$ pairwise exchanges. We take $N=1000$ agents and total money $M=N$,  initially distributed over all the agents uniformly. The steady state distribution is measured over $10^3$ Monte Carlo time steps  after relaxing  $10^6$  Monte Carlo time steps for equilibration.

\subsection{Uniform saving income exchange models and inequality indices}
\label{cc-model}
In this model, we consider a conserved system where total money $M$ and total agents $N$ are fixed. Each agent i possesses money $m_i(t)$ at a time t and in any interaction, a pair of agents i and j exchange their money  such that their total money is  conserved. For fixed saving propensity $\lambda$ of the agents,  the exchange of money between two randomly chosen pairs can be expressed as
\begin{equation}
\begin{split}
    m_i(t+1) &= \lambda m_i(t)+\epsilon_{ij}((1-\lambda)(m_i(t)+m_j(t))) \\
    m_j(t+1) &= \lambda m_j(t)+(1-\epsilon_{ij})((1-\lambda)(m_i(t)+m_j(t)))
    \end{split}
    \label{eqn1}
 \end{equation}   
where  $0 \leq \epsilon_{ij} \leq 1 $ is a random fraction varying in every interaction.

\begin{figure}[!tbh]
    \centering
    \includegraphics[width=1.0\textwidth]{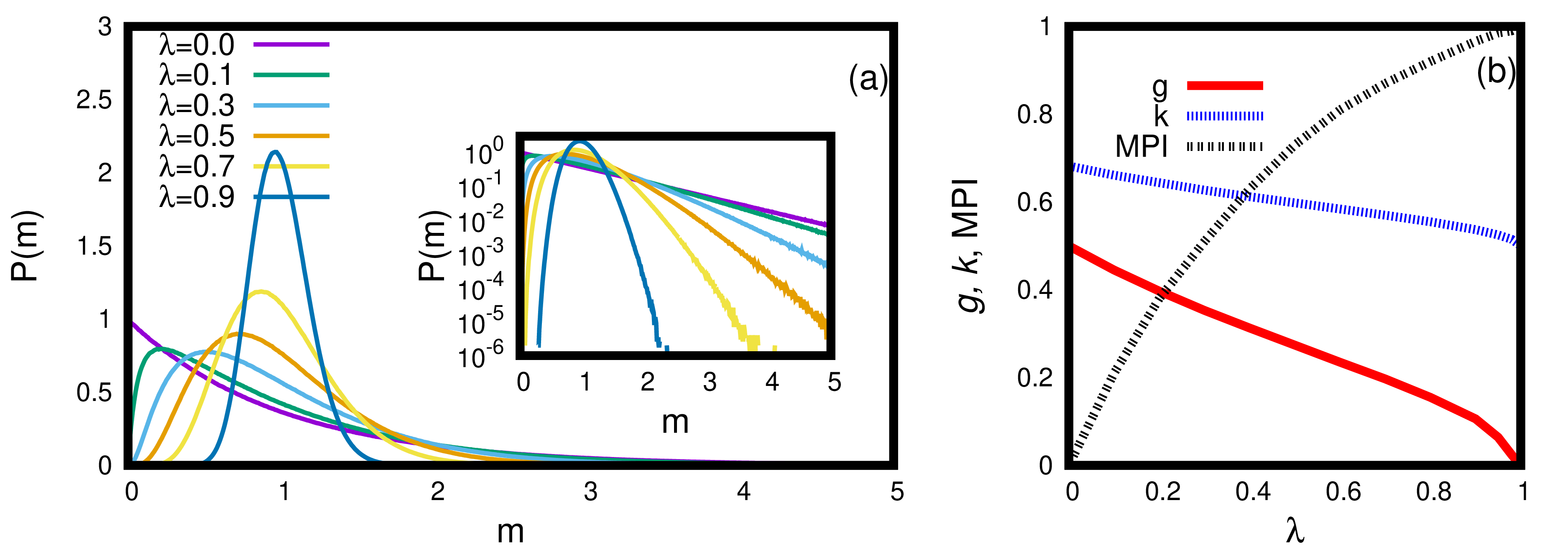}
    \caption{The pure kinetic exchange dynamics: (a) The steady state income distributions P(m) for different saving propensity $\lambda$ are shown in the plot. (Inset) The same distributions are shown in semi-log scale indicating an exponential nature of the tail end of the  distributions.  (b) The variation of Kolkata index ($k$), Gini index ($g$) and Most Probable Income (MPI) are shown against fixed saving propensity $\lambda$ (maximum value of $\lambda$ is $1_{-}$).}
    \label{fig:2}
\end{figure}

The steady state income distribution $P(m)$ for fixed saving propensity $\lambda$ is shown in Fig. \ref{fig:2}(a). For $\lambda=0$,  the steady state follows a Gibbs distribution and Most Probable Income (MPI) distribution is at MPI= 0. MPI per agent shifts from m = 0 to m = 1 as $\lambda\to1$.  Furthermore, the plot in semi-log (as shown in Fig. \ref{fig:2}(a) inset) indicates an exponential nature for the tail end of the distribution.

 In plot \ref{fig:2}(b), we show the variation of the Kolkata index ($k$), the Gini index ($g$) and the Most Probable Income (MPI) against saving propensity $\lambda$. The Gini coefficient value  diminishes  from $0.5$ to $0$ as $\lambda$ approaches from $0$ to $1$. Similarly the $k$-index value reduces from $0.68$ to $0.5$ as $\lambda$ approaches from $0$ to $1$.

\subsection{Self-organized minimum poverty level model and inequality indices}
\label{soc-model}
\begin{figure}[!tbh]
    \centering
    \includegraphics[width=1.0\textwidth]{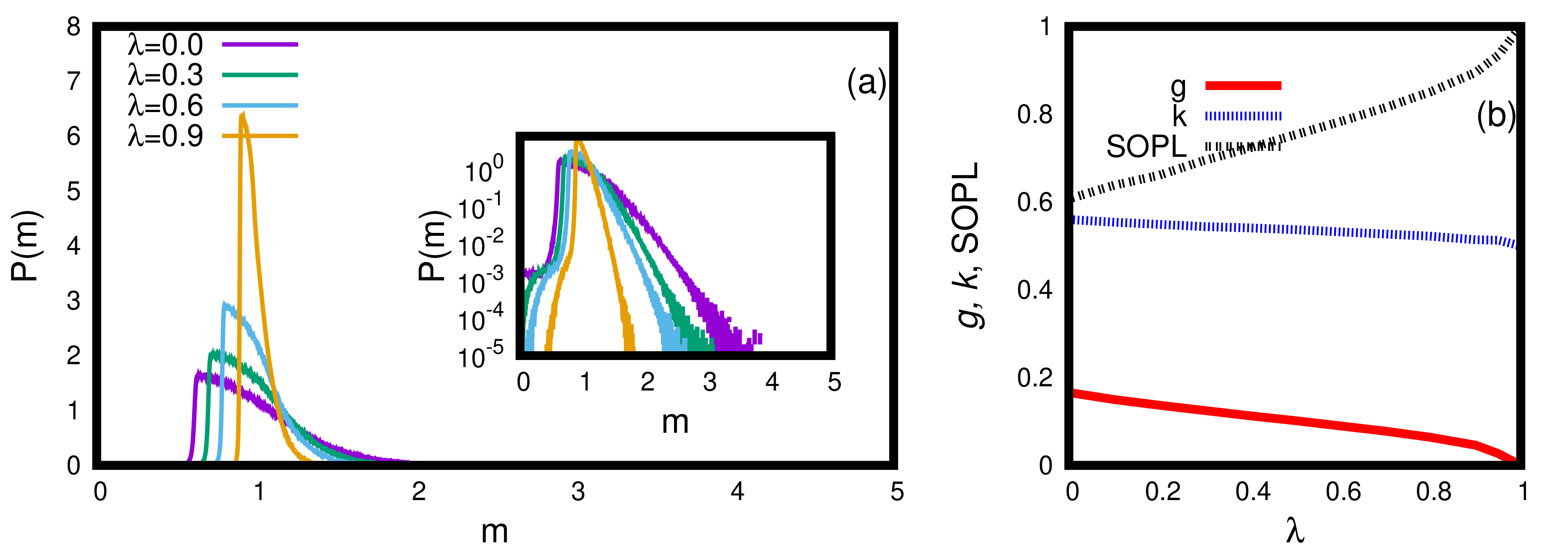}
    \caption{Self-Organized Poverty Level model: (a) Steady state income distribution $P(m)$ for fixed saving propensity $\lambda$ are shown.  (Inset) The same distributions are shown in semi-log scale  indicating an exponential nature of the tail end of the  distributions. (b) The variation of the Kolkata index ($k$), the Gini index ($g$) and location of the  Self-Organized Poverty Level (SOPL) are shown against fixed saving propensity $\lambda$ (maximum value of $\lambda = 1_{-}$).}
    \label{fig:3}
\end{figure}

Here we consider a model where one of the agents in the chosen pair is necessarily the poorest at
that point of time and the other one is randomly chosen from the rest.
Here we vary $\lambda$ for values other than 0, as shown in Fig. \ref{fig:3}. 
The exchange of money will follow
the same rule as described by Eqn. \ref{eqn1}. An important observation that ensues 
from the $\lambda\neq 0$ curves is that  
the MPI displaces to the right of the spectrum and 
shifts close to $m =1$ (resembling the nature of $P(m)$ against $m$ in Fig. \ref{fig:2}). Here we also observe that the SOPL (below which $P(m)=0$) increases with increasing values of $\lambda$.

In Fig. \ref{fig:3}(a), the steady state income distribution $P(m)$ for different  $\lambda$ values are shown and   the same distributions are shown in semi-log scale in the inset indicating an exponential nature
of the tail end of the  distributions. In Fig. \ref{fig:3}(b), the variation of Kolkata index ($k$), Gini index ($g$) and Self-Organized Poverty-Line or SOPL are shown against saving propensity $\lambda$. The figure indicates that inequality of the distribution diminishing as $\lambda\to1$  and also Self-Organized Poverty-Line or SOPL is rising to 1 as $\lambda\to1$.

\subsection{Indices for pure kinetic exchange model with two choices of  $\lambda$}
\label{cc-two}

A more realistic saving behavior in a society would definitely mean that $\lambda$ should vary over time 
as it is heavily dependent on an individual's saving interests. Thus, a randomness attached to the 
saving propensity in the form of a probability factor $\rho$ can act as an additional parameter in determining
the nature of the wealth distribution curve. With this in mind, we consider exchange dynamics similar to that described in subsection \ref{cc-model},  but with a slight modification of Eqn. (\ref{eqn1}) as follows: \\
 \begin{equation}
 \begin{split}
  m_i (t + 1) = \lambda_{\rho} m_i (t) +\epsilon_{ij} ((1 - \lambda_{\rho})(m_i (t) + m_j (t)))\\
  m_j (t + 1) = \lambda_{\rho} m_j (t) + (1 - \epsilon_{ij} )((1 - \lambda_{\rho})(m_i (t) + m_j (t)))
  \end{split}
  \label{eqn2}
 \end{equation}\noindent using $\lambda_{\rho} = 1$ with probability  $\rho$ and  $\lambda_{\rho} = 0$ with probability $1-\rho$;  $\rho <1$.
Henceforth we will refer to $\lambda_{\rho}$ as $\lambda$, without loss of generality
and provided Eqn. \ref{eqn2} is satisfied.

Eqn. \ref{eqn2} suggests that each agent has two choices of $\lambda$ over time. In our study, the agents can take the saving propensity either 1 (with probability $\rho$) or 0 (with probability $1-\rho$) over time.
For $\lambda$ strictly equal to 1, the exchange dynamics stops and hence values of $\lambda$ infinitesimally close to 1
are considered for simulations.

\begin{figure}[!tbh]
    \centering
    \includegraphics[width=1.0\textwidth]{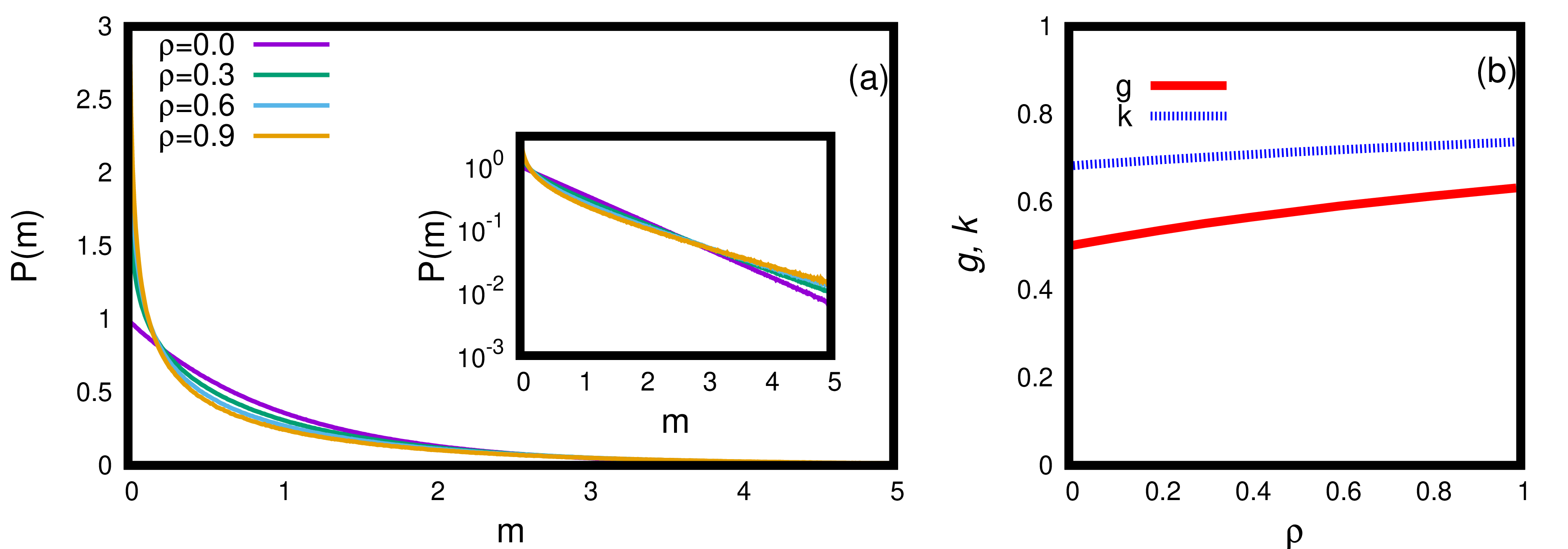}
    \caption{The pure kinetic exchange model of two choices of  $\lambda$. (a) Steady state income distribution $P(m)$ for fixed saving propensity $\lambda$ are shown.  (Inset) The same distributions are shown in semi-log scale  indicating an exponential nature of the tail end of the  distributions.  (b) The variation of Kolkata index ($k$) and Gini index ($g$) are shown against  the probability for taking $\lambda=1$ i.e. $\rho$ (maximum value of $\rho = 1_{-}$).}
    \label{fig:4}
\end{figure}

In Fig. \ref{fig:4}(a), the steady state income distribution $P(m)$ is shown for different  probability $\rho$. We
observe that the most probable income (MPI) occurs at $m=0$, and the semi-log plots of the
distributions indicate the exponential nature of the tail end of the distribution (see inset of Fig. \ref{fig:4}(a)).  The  Kolkata index ($k$) and Gini index ($g$) rise slowly against $\rho$ (see Fig. \ref{fig:4}(b)).

\subsection{Self-organized minimum poverty level model: Indices for  two choices of  $\lambda$}
\label{soc-two}
\begin{figure}[!tbh]
    \centering
    \includegraphics[width=1.0\textwidth]{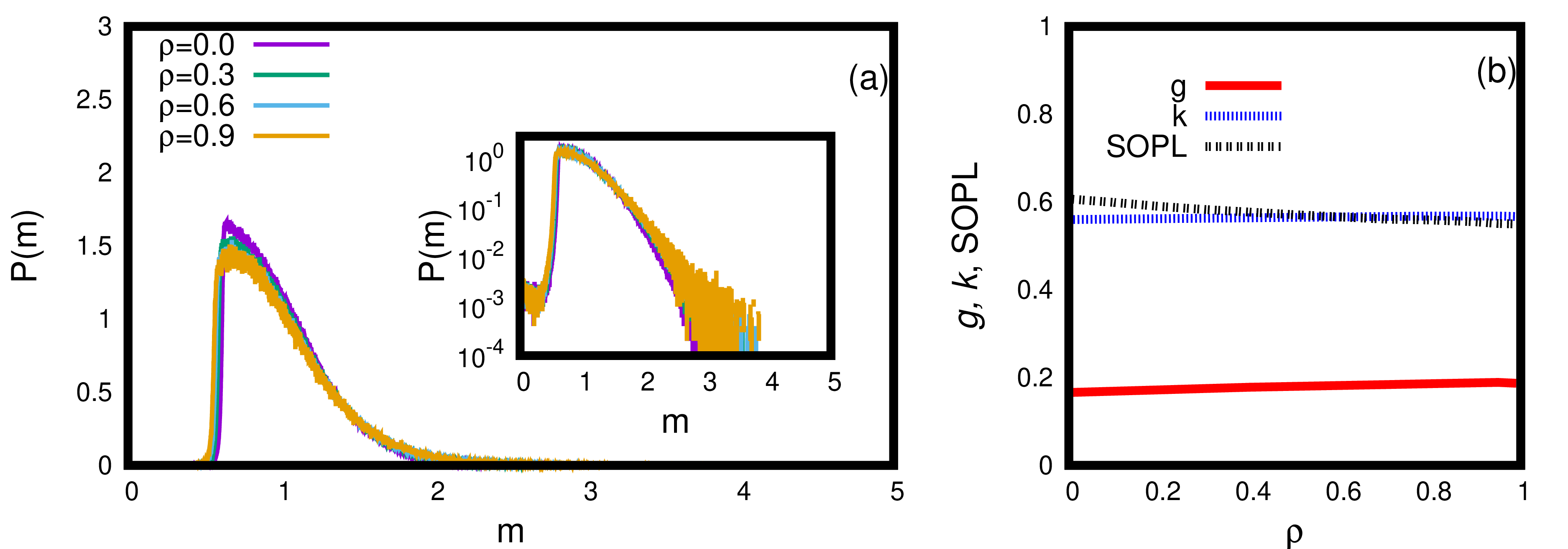}
    \caption{A self-organized minimum poverty level} model for two choices of saving propensity: (a) The steady state income distribution $P(m)$ for fixed saving propensity $\lambda$ is shown.  (Inset) The same distributions are shown in semi-log scale  indicating an exponential nature of the tail end of the  distributions. (b) The variation of the Kolkata index ($k$), the Gini index ($g$) and the Self-Organized Poverty-Line or SOPL are shown against fixed saving propensity $\lambda$. A very slow increasing trend of the inequality indices and a slow decreasing trend of SOPL against $\rho$ can be observed here (maximum value of $\rho = 1_{-}$).
    \label{fig:5}
\end{figure}

As before, we consider the same dynamics as described in subsection \ref{soc-model} but the difference is that each agent has two choices of $\lambda$ over time. In our study, the agents can take the saving propensity either 1 (with probability $\rho$) or $0$ (with probability $1-\rho$) over time and we assume that the poorest agent trades with any of the agents in the rest of the population, which not only leads to a finite MPI but also a self organization in the system leading to an SOPL.  In Fig. \ref{fig:5}(a), the steady state income distribution $P(m)$ is shown for different probability $\rho$.  The  semi-log plots of the distributions indicate the exponential nature of the distribution (see inset of Fig. \ref{fig:5}(a)).  The variation of Kolkata index ($k$) and Gini index ($g$) and Self-Organized Poverty-Line or SOPL are shown against $\rho$ in  Fig. \ref{fig:5}(b). A very slow increasing trend of the inequality indices  with $\rho$ can be observed here. Also the SOPL of the distribution slowly decreases against $\rho$.

\section{Summary \& Discussion}
\label{conclusions}
Income and wealth inequality in human societies
has been ubiquitous over the history of mankind,
very well recognized and has long been studied
by philosophers and economists (see e.g., \cite{champernowne1998economic,piketty2013capital}
for some recent discussions by economists).
Recently, physicists have also joined the
investigations on income and wealth inequalities
(see e.g, \cite{chakrabarti2013econophysics,pareschi2013interacting,sen2014sociophysics} for some detailed discussions and \cite{ludwig2021physics}
for an interesting application in the context of
US economic inequalities during 1983-2018) 
essentially using kinetic theory or specifically
kinetic exchange models (the key theme of
this special issue).

In this paper, we have numerically studied the variations
of income or wealth distribution $P(m)$ generated
from the kinetic exchange models described earlier in section \ref{results}
and extracted the variations
in the values of Gini  $g$ and Kolkata $k$ indices with  the saving  propensity $\lambda$ of the agents, with two different kinds of trade or (kinetic) exchange dynamics. 
Along with the  nature of
steady state distributions $P(m)$, we have also studied the variations in the values of the inequality indices ($g$ and $k$) and the location of the Most Probable Income  (MPI) or the Self-Organized Poverty Level (SOPL, if any) of income or wealth as the saving propensity $\lambda$ of the agents (fixed over time) and as the time fraction $\rho$ of choosing the saving propensity $\lambda = 1$ over the other choice $\lambda = 0$. Although two of the versions of the
models (discussed in subsections \ref{cc-model} and \ref{soc-model}) were
addressed in earlier studies, the results for the variations
of Gini ($g$)  and  Kolkata ($k$) indices with the
saving propensity ($\lambda$) of the agents are new. The decrease in the inequality indices re-establish 
the models as sustainable market models where sheer self-interest in saving a 
portion of the money traded by each individual locally can cause a global 
peaked wealth distribution, where most agents prosper with finite money. We note in subsection \ref{cc-two}, where pair of traders are chosen randomly, and for the case discussed in subsection \ref{soc-two}, where the poorest agent is one of the partners in trade that both $g$ and $k$ increase (see Fig \ref{fig:4}b) and SOPL decreases weakly (see Fig. \ref{fig:5}b). These are probably justified because as $\rho$ increases, interactions with fully corrupt  persons (having $\lambda=1$, who accumulate money in each trade and spend nothing) become more frequent. 
However, an important observation is that after introducing the constraint of selecting at least one agent
with minimum money, an SOPL emerges (Fig. 5a) and so does an MPI
replacing the $\lambda = 0$ exponential to a Gamma-like distribution for all
values of $\rho$. Hence an incentivized trading process for the poor, with
just offering them greater opportunity to trade without supporting
them with extra wealth, works greatly in raising the minimum wealth
possessed by a population to a threshold money.

As shown in Figs. 2-5, the most-probable income
or MPI (where $P(m)$ is highest) or the
self-organized poverty level, the SOPL (below
which $P(m) = 0$ and usually the MPI coincides
with the SOPL) increases with increasing saving
propensity or $\lambda$.  Generally speaking, in all these fixed saving
propensity cases (see Figs. \ref{fig:2} and \ref{fig:3}), the
income or wealth inequalities, as measured by the
index values of Gini $g$ and Kolkata $k$\ (= 0.5 and
$\simeq$ 0.68 respectively, in the pure
kinetic exchange or Gibbs case, discussed
analytically in the Introduction) decreases with
increasing saving propensity ($\lambda$) of the
agents.

Economic policy prescriptions naturally aim for
reduced economic inequalities (see e.g., \cite{venkatasubramanian2017much} for
some discussions on the limits of equality imposed
by thermodynamic and physical considerations).
In this context, as our study indicates, reduction in the values of both
the Gini and the Kolkata indices and increase in the value
of the self-organized poverty level with increasing
saving propensities of the agents, look quite encouraging. In particular,
it should help in building and studying more specific
kinetic exchange models with useful economic
policy implications.

\section*{acknowledgments}
BKC is thankful to the
Indian National Science Academy for their
Senior Scientist Research Grant. BJ is
grateful  to the Saha Institute of Nuclear
Physics  for the award of their
Undergraduate Associateship.

\bibliographystyle{unsrt}
\bibliography{ref.bib}

\end{document}